\documentclass[aps,prd,twocolumn,groupedaddress, showpacs, letterpaper]{revtex4}
\usepackage{graphics}
\usepackage{subfig}
\usepackage{color, amsmath, epsfig, amssymb,mathrsfs}

\begin{document}

\preprint{Amitai's paper}

\title{Gravitational Lensing of Stars Orbiting Sgr A* as a Probe of the Black Hole Metric in the Galactic Center}

\author{Amitai Y. Bin-Nun}
\email[]{binnun@sas.upenn.edu}
\affiliation{Department of Physics and Astronomy, University of Pennsylvania}

\date{\today}

\begin{abstract}
We show that a possible astrophysical experiment, detection of lensed images of stars orbiting close to Sgr A*, can provide insight into the form of the metric around a black hole. We model Sgr A* as a black hole and add in a $\varpropto \frac{1}{r^2}$ term to the Schwarzschild metric near the black hole. We then attempt to determine the effect of this extra term on the properties of the secondary images of the S stars in the Galactic Center. When the $\frac{1}{r^2}$ term is positive, this represents a Reissner-Nordstrom (RN) metric, and we show that the there is little observational difference between a Schwarzschild black hole and a RN black hole, leading to the conclusion that secondary images may not be a useful probe of electrical charge in black holes. A negative value for the $\frac{1}{r^2}$ term can enter through modified gravity scenarios. Although physically unlikely to apply in the case of a large black hole, the Randall-Sundrum II braneworld scenario admits a metric of this form, known as tidal Reissner-Nordstrom (TRN) metric. We use values of tidal charge ($Q$ in $\frac{Q}{r^2}$) ranging from $-1.6 M^2$ to $0.4 M^2$. A negative value of $Q$ enhances the brightness of images at all times and creates an increase in brightness of up to 0.4 magnitudes for the secondary image of the star S2 at periapse. We show that for other stars with brighter secondary images and positions  more aligned with the optic axis, using the Tidal Reissner-Nordstrom metric with negative $Q$ enhances the images as well, but the effect is less pronounced. This effect is related to the increase in the size of the photon sphere, and therefore, should be noticeable in other metrics with a similar effect on the photon sphere. With the next generation of instruments and increased knowledge of radiation from Sgr A*, using properties of secondary images to place constraints on the size of the $\frac{1}{r^2}$ term. This knowledge will be useful in constraining any modified gravity theory that adds a similar term into the strong field near a black hole. 
\end{abstract}


\maketitle

\section{Introduction}

Gravitational lensing provided the first experimental verification of general relativity (GR) through observations of starlight bending around the Sun during an eclipse in 1919 \cite{schneider, 1919eclipse} and continues to be a major source of insight into gravitation \cite{LensingReview, schneider, jainreview}. Excitingly, increasingly precise observations of the compact radio source Sgr A* at the Galactic center and its surrounding stars have given us very high confidence that only a very gross deviation from GR could allow for the absence of a black hole there \cite{NatureHorizon,smbhdetails, ghez2008, propSgrA, propSgrA2}.  According to \cite{propSgrA}, the black hole is estimated to have a mass of about $4.31 \times 10^6 M_{\odot}$ and a distance of about 8.33 kpc from Earth. Black holes are unique laboratories for gravitational lensing because their compactness allows light to closely approach the photon sphere and its path will bend significantly there due to gravity. Most studies of gravitational lensing \cite{petters, petterskeeton} are in the weak deflection limit, when the point of closest approach of any lensed photons is far from the lensing mass. This allows for the simple expression of the bending angle as $\alpha = \frac{4M}{r_0}$, where $r_0$ is the point of closest approach of the null geodesic (this paper uses geometric units $G=c=1$). This expression is valid in the limit $\frac{M}{r_0} \ll 1$. When a photon closely approaches a  black hole's photon sphere, the weak deflection limit does not hold and using this approximation for the bending angle leads to inaccurate results. For a spherically symmetric, static metric with line element

\begin{equation}
ds^2= -A(r)dt^2 + B(r) dr^2 +C(r) r^2 d\Omega^2
\label{metric}
\end{equation}
the bending angle is an elliptic integral based on the functions of the metric \cite{VE2002} and is
\begin{eqnarray}
\nonumber \alpha (r_0) &=& 2 {\int_{r_0}}^{\infty}\left(\frac{B(r)}{C(r)}\right)^{1/2}
                       \left[(\frac{r}{r_0})^2\frac{C(r)}{C(r_0)}\frac{A(r_0)}{A(r)}-1\right]^{-1/2} \\ 
& \times & \frac{dr}{r}- \pi
      \label{bending}
\end{eqnarray}

 If the point of closest approach is very close to the photon sphere, the strong deflection limit approximation of this integral \cite{bozza2002, bozza2007} can be used. In all cases, a full numerical treatment of the bending angle can be used \cite{VE2000}, and it should be used for studies, such as this one, where neither the strong nor weak deflection limits are satisfied. This is further explained in Sec. \ref{sec:lensing}. Large bending angles yield interesting results: when photons approach close enough to the photon sphere, they can loop around the lens before reaching the observer and produce an infinite sequence of images on both sides of the optic axis \cite{darwin, VE2000, bozzaetal, bozza2002, V2009}. The properties of these images are sensitive to the form of the black hole metric \cite{eiroad, whisker2005, me, mmreview, whiskerthesis} with \cite{eiroarn} studying the effect of the RN metric. There has been further study of an astrophysical scenario which utilizes lensing with a large bending angle. S2 was the first star that was examined for a ``retrolensing" \cite{depaolis} effect (lensing with a bending angle of $ \alpha \approx \pi$).  By using orbital parameters of the S stars in the Galactic center provided by \cite{propSgrA}, \cite{bozzas2, bozzas2revised,bozzamancini2009} study the properties of secondary images of those S stars. This study showed 9 stars with secondary images with a maximum brightness in the K-band brighter than $m_K=30$, a cutoff that maximizes observational prospects.
 
In this paper, we use a black hole metric with an additional term that is $\varpropto \frac{1}{r^2}$. The metric is
\begin{equation}
ds^2 = -(1-\frac{2M}{r} + \frac{Q}{r^2})dt^2 + (1-\frac{2M}{r}+  \frac{Q}{r^2})^{-1} dr^2 +r^2 d\Omega^2
\label{eq:trnmetric}
\end{equation}
with $ Q$ a free parameters often expressed by $q \equiv \frac{Q}{4M^2}$. When $q$ is positive, this represents the Reissner-Nordstrom solution for a charged black hole. Static black holes with a large amount of electric charge are not expected to exist in nature, and the existence of rotating, charged black holes is controversial \cite{punsly}. In addition, the amount of charge is limited to $Q < M^2$ or $q < 0.25$ because the saturation of this bound would lead to a naked singularity and the violation of cosmic censorship \cite{waldcensorship}. However, since the calculation of the properties of secondary images of S stars has only been done with a Schwarzschild metric, it is useful to examine such a fundamental case. We have also found it useful to explore negative values of $q$. There are only weak constraints on a lower bound for $q$ that come from studies of neutron star binary systems \cite{trnconstraints}. Using a $\frac{1}{r^2}$ term is interesting because it is motivated by alternative gravity frameworks, particularly braneworld theories that construct gravity as a higher-dimensional theory. As these theories usually predict a correction that strengthens gravity, this would correspond to a negative value for $q$. For any non-trivial value of $q$, observational constrains from Solar System observations \cite{dmptidalrn} disallow the $\frac{1}{r^2}$  term from applying anywhere but near the black hole. Strengthening gravity using a negative value for $q$ yields a brighter secondary image; therefore, observations of these secondary images can place constraints on the size of the $\frac{1}{r^2}$ term near the black hole.   

An extra $\frac{1}{r^2}$  term in the metric comes directly from a proposed black hole metric in the Randall-Sundrum II theory. The Randall-Sundrum I model \cite{rs1} is a braneworld scenario inspired by heterotic M-theory \cite{ovrut1, ovrut2, ovrut3}, a 5 dimensional effective framework that arises from the dimensional reduction of 11-dimensional Harova-Witten theory on a Calabi-Yau manifold with the imposition of $S^1/Z_2$  symmetry \cite{witten1}. Six dimensions are compactified, making gravity effectively 5 dimensional. The Randall-Sundrum II scenario is this model with the second brane taken to infinity \cite{rs2, MaartensReview}. It is not clear whether a static black hole solution exists for a RS II braneworld. Although \cite{emparan} has shown the existence  of a static black hole in a 2+1 brane setup, it is unclear what the solution is and whether there even is a static black hole solution in the  3 + 1 brane scenario \cite{kanti, creek, shiromizu2, wiseman, wiseman2}. Attempts to study the problem have yielded several possible black hole metrics \cite{dmptidalrn, whisker2005, blackstring, casadio, gtmetric}. The effect of these metrics on lensing has been examined in several studies \cite{eiroad, whisker2005, me, petterskeeton, mmreview}. For a supermassive black hole such as Sgr A*, the Garriga-Tanaka \cite{gtmetric} and ``black string"  \cite{blackstring} metrics will not show any results \cite{me} and this paper aims to show that a modification of the metric near the horizon can affect lensing observables. One particular metric that has been studied in connection with lensing by the black hole at Sgr A* is the ``tidal" Reissner-Nordstrom metric \cite{dmptidalrn, eiroad, whiskerthesis, me} which is of the form of Eq. (\ref{eq:trnmetric}). We therefore use it as an example of a metric with a $\frac{1}{r^2}$ potential term. There are several possible objections to using the TRN metric for a supermassive black hole: Studies suggest that supermassive black holes in the braneworld should have induced metrics that are no more than negligibly different than the Schwarzschild metric \cite{wiseman}. In addition, the TRN metric has no known completion in the bulk \cite{whiskerthesis} and there are likely naked singularities in the bulk for the TRN metric \cite{trnpath}. However, we will study lensing properties of the TRN metric to gain understanding of the effects of adding a $\varpropto \frac{1}{r^2}$ term in the potential, whether it comes from the braneworld scenario or any other gravitational framework.

In Sec. II, we discuss S stars near the Galactic center. Now that their orbital parameters are well known from decades of observation \cite{propSgrA}, we show how to reconstruct the orbits of these stars and characterize the variables in the Ohanian lens equation \cite{bozzalens, ohanian} in terms of the star's orbital parameters. Although current uncertainties in lensing parameters can interfere with the test we are proposing, future instruments and observations will further constrain these orbital parameters. In Sec. III, we use the orbital parameters to construct a light curve for the secondary images of S2, S6, and S14. For these stars, we show the light curve of the secondary image when assuming a Schwarzschild spacetime, a TRN spacetime with $q= -1.6$, and, for the case of S2, an extremal RN spacetime with $q= 0.25$. We show that for high enough values of the tidal charge parameter $q$, an appreciable difference appears for these light curves in the TRN spacetime. We also show the relevant properties, such as image magnitude and image position at peak brightness for several values of $q$. In Sec. IV, we discuss the observational prospects for these images as well as the observing the difference in brightness for these images. We conclude that observation of these images is possible, as is observing the difference in image properties due to a modified gravity theory.

\section{\label{sec:lensing}Lensing of S Stars}

There is a large stellar population in the central parsec of the Milky Way. Most stars are old red main sequence stars, but there are many young Wolf-Rayet stars which present challenges to theories of star formation \cite{gcstellar, ghez2003}. Some of these young stars orbit close to the black hole at Sgr A*, labelled S stars, have been observed carefully. Their orbits can be reconstructed from orbital parameters published in \cite{propSgrA}. For the purposes of this paper, at each point in time, we treat the star as a source being lensed by the black hole. To study the affect of the metric near the black hole, we are interested in the light that passes across the optic axis (line connecting the lens and the observer), forming a secondary image \cite{schneider, petters}. Using orbital parameters and the intrinsic brightness of the stars in the K-band, \cite{bozzamancini2009, bozzas2, bozzas2revised} calculated the position and magnitude of the secondary images, assuming a Schwarzschild metric. Observationally, the most promising images come from the stars S6 and S27, which have the brightest predicted secondary images with $m_K= 20.8$ and $22.4$ respectively. The peak brightness occurs when each star is at periapse. Although it will not get as bright, the periapse of S2 will occur relatively soon in 2018 and its secondary image will be at peak brightness then, with predicted $m_K= 26.8$. Although S27 is brighter, we will instead study S14 because its peak brightness comes at an earlier date than the peak of S27. In addition, the secondary image of S14 is a better candidate for differentiating between a Schwarzschild spacetime and an alternative one, as will be explained in Sec. \ref{sec:results}.

In calculating the positions of the secondary image, we use the improved Ohanian lens equation \cite{ohanian, bozzalens}, apply the small angle approximation, and throw out negligible terms, leaving 
\begin{equation}
\gamma= \alpha(\theta)-\frac{D_L}{D_{LS}}\theta
\label{eq:ohanian}
\end{equation}
where $\gamma$ is the angle between the optic axis and the line connecting the source to the lens, $\theta$ is the image position (to the observer), $D_L$ is the constant distance from the observer to the lens (in this case, the distance between us and Sgr A*) and $D_{LS}$ is the distance, which varies over time, between the lens and the source star. The anomaly angle from the periapse ($\phi$) is determined by the initial conditions and differential equation:

\begin{equation}
\frac{[a(1-e^2)^{\frac{3}{2}}}{\sqrt{GM_{enc}}(1+e \cos \phi)^2}\dot{\phi}^2=1
\end{equation}
where $GM_{enc}$ is the mass enclosed within the orbit and $e$ is the eccentricity of the orbit. $GM_{enc}$ can be calculated in relation to the orbital period $P$ and semi-major axis $a$ using Kepler's Third Law
\begin{equation}
GM_{enc}= 4 \pi^2 \frac{a^3}{P^2}
\end{equation} 
The relationship between $\gamma$ and $D_{LS}$ and $\phi$ is given by:

\begin{eqnarray}
D_{LS} &=& \frac{a(1-e^2)}{1+e \cos \theta} \\
\gamma &=& \arccos[ \sin (\phi + \omega)\sin i]
\end{eqnarray}

In Eq. (\ref{eq:ohanian}), $\alpha(\theta)$ is the bending angle as a function of $\theta$. As noted in \cite{bozzamancini2009}, the Ohanian lens equation is accurate to better than $10^{-6}$. Applying the small-angle approximations and throwing out the $\theta$ term to obtain Eq. (\ref{eq:ohanian}) affects outcomes by less than that. For a given source position, we numerically solve Eq. (\ref{eq:ohanian}) which yields the position of the secondary image. The image magnification is \cite{bozzamancini2009}:

\begin{equation}
\mu= \frac{D_L^2}{D_{LS}^2}\frac{\sin \theta}{\frac{ d\gamma}{d \theta} \sin \gamma}
\label{eq:mag}
\end{equation}
Analytical formulas for image position and magnification are given by \cite{bozzas2}. However, the analytical formulas are not accurate because the bending angle in this astrophysical scenario often lies neither in the strong or weak deflection limit. Therefore, numerical evaluation of the image positions and magnifications are necessary. To calculate this, we must numerically evaluate Eq. (\ref{bending}). Writing down the equations of motion for photons in this spacetime 
\begin{equation}
g_{\mu\nu}\dot x^\mu\dot x^\nu=0 
\end{equation}
the coordinates $\phi$ and $t$ are cyclic, leading to the conserved quantities
\begin{eqnarray}
E &=& B(r) \dot{t}\\
J &=& C(r) r^2 \dot{\phi}
\end{eqnarray}
which characterize the motion of null geodesics throughout this spacetime \cite{weinberg, bozza2007}. Importantly, we can create a relationship between the coordinate of closest approach ($r_0$) and the angular position of images \cite{VE2000}:

\begin{equation}
J= D_L \sin \theta = r_0 \sqrt{\frac{C(r_0)}{B(r_0)}}
\label{eq:theta}
\end{equation}
we can then take the derivative of $\alpha$ with respect to $\theta$
\begin{equation}
\frac{ d\alpha(r_0)}{d\theta}= \frac{ d\alpha(r_0)}{dr_0} \frac{ dr_0}{d\theta}
\end{equation}
which is necessary to evaluate Eq. (\ref{eq:mag}). This method is described in greater detail in \cite{me, VE2000}. 

\begin{table*}[t]

\begin{tabular}{c c c c c c c c c}
\hline
\hline
  Star & $a ["]$ & $e$ & $i[^{\circ}]$ & $\Omega[^{\circ}]$ & $\omega[^{\circ}]$ & $t_P$[yr]& T[yr]&K \\
\hline
S2 & 0.123 $\pm$ 0.001 & 0.88 $\pm$ 0.003 & 135.25 $\pm$ 0.47 & 225.39 $\pm$ 0.84 & 63.56 $\pm$ 0.84 & 2002.32 $\pm$ 0.01& 15.8 $\pm$ 0.11& 14  \\
S14 & 0.256 $\pm$ 0.01 & 0.963 $\pm$ 0.006 & 99.4 $\pm$ 1. & 227.74 $\pm$ 0.7 & 339 $\pm$ 1.6 & 2000.07 $\pm$ 0.06& 47.3 $\pm$ 2.9& 15.7  \\
S6 & 0.436 $\pm$ 0.153 & 0.886 $\pm$ 0.0026 & 86.44 $\pm$ 0.59 & 83.46 $\pm$ 0.69 & 129.5 $\pm$ 3.1 & 2063 $\pm$ 21 & 105 $\pm$ 34& 15.4  \\
\hline
\end{tabular}
\caption{\label{table:stars}Orbital Parameters of the S Stars examined in This Letter: $a$ is the semimajor axis, $e$ is the eccentricity, $i$ is the inclination of the normal of the Orbit with respect to the line of sight, $\Omega$ is the position angle fo the ascending node, $\omega$ is the periapse anomaly with respect to the ascending node, $t_P$ is the epoch of either the last or next periapse, $T$ is the orbital period, and $K$ is the apparent magnitude in the $K$ band (data taken from \cite{propSgrA})}
\end{table*}

\section{\label{sec:results} Results}

\begin{figure}[h]
 \begin{center}
\includegraphics[width=0.5 \textwidth]{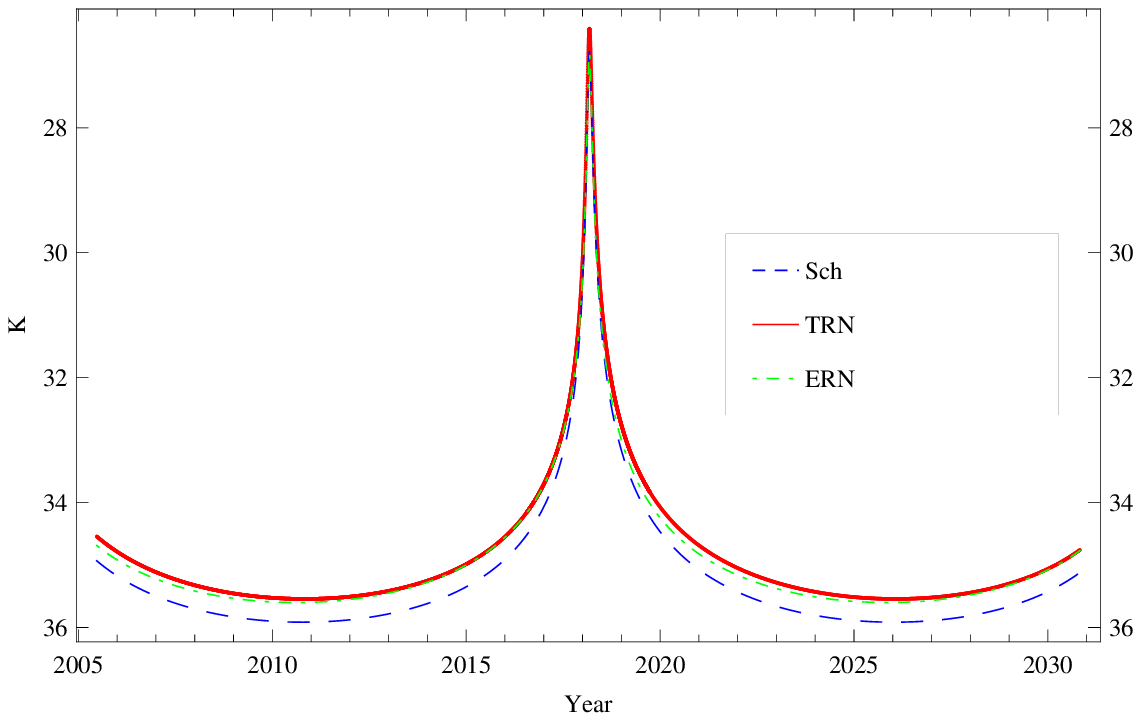}
\includegraphics[width=0.5 \textwidth]{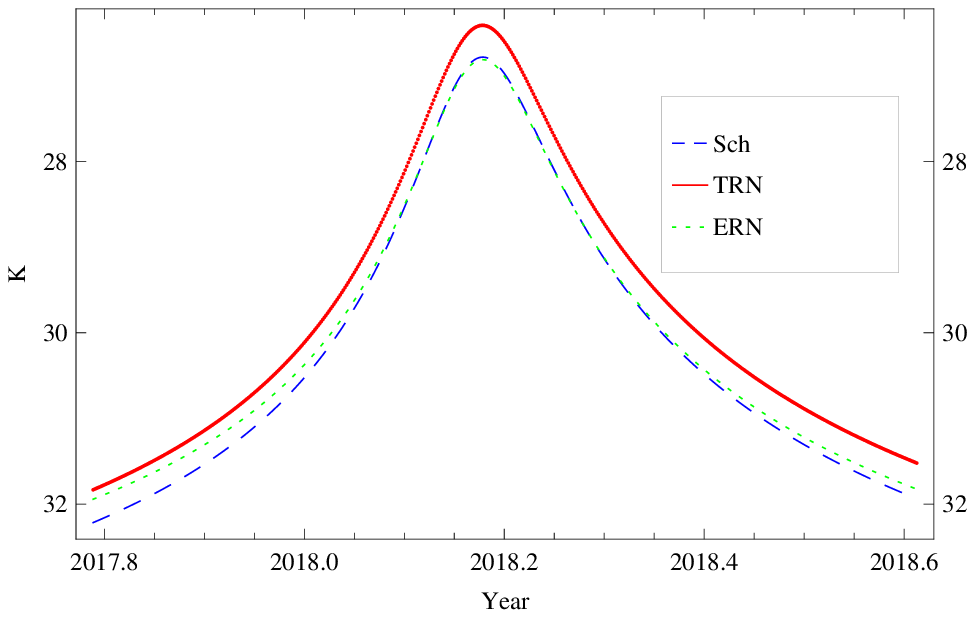}
 \caption{The light curve for the secondary image of S2 is dependent on the choice of metric in the strong field. The top figure contains light curves for an entire period of the orbit of S2. The bottom image contains light curves in the year around periapse and peak brightness.}
 \label{fig:s2curve}
 \end{center}
 \end{figure}

Using a variation of this algorithm and the values in Table \ref{table:stars}, Bozza calculated the light curve of the secondary image of S2. He found a peak brightness of $m_K=26.8$. In Fig. \ref{fig:s2curve} we have produced a graph that compares the light curve  for the secondary image of S2 using the Schwarzschild metric with light curves calculated using a metric with a $\frac{q}{r^2}$ term. In Fig. \ref{fig:s2curve}, we calculate the light curve for $Q= M^2$ in Eq. (\ref{eq:trnmetric}), which is equivalent to $q= 0.25$. This represents an extremal RN black hole- this or any higher value of $q$ would result in a naked singularity and is expected to be non-physical. We also calculate the light curve for the S2 star when we set $q=-1.6$ and plot it on the same graph. 

\begin{figure}[h]
 \begin{center}
\includegraphics[width=0.5 \textwidth]{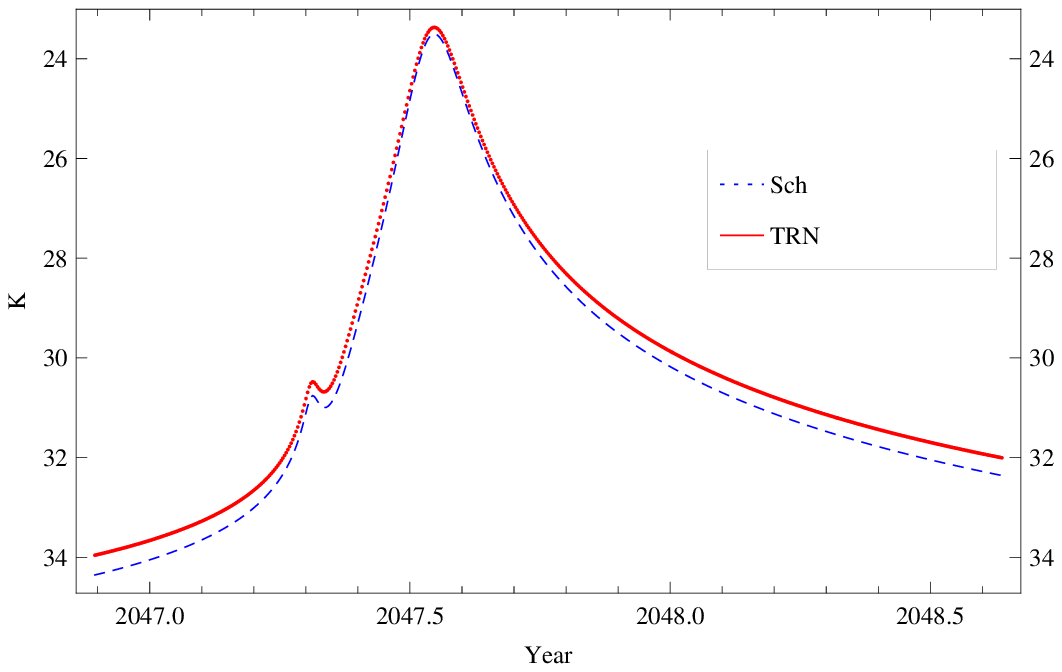}
\includegraphics[width=0.5 \textwidth]{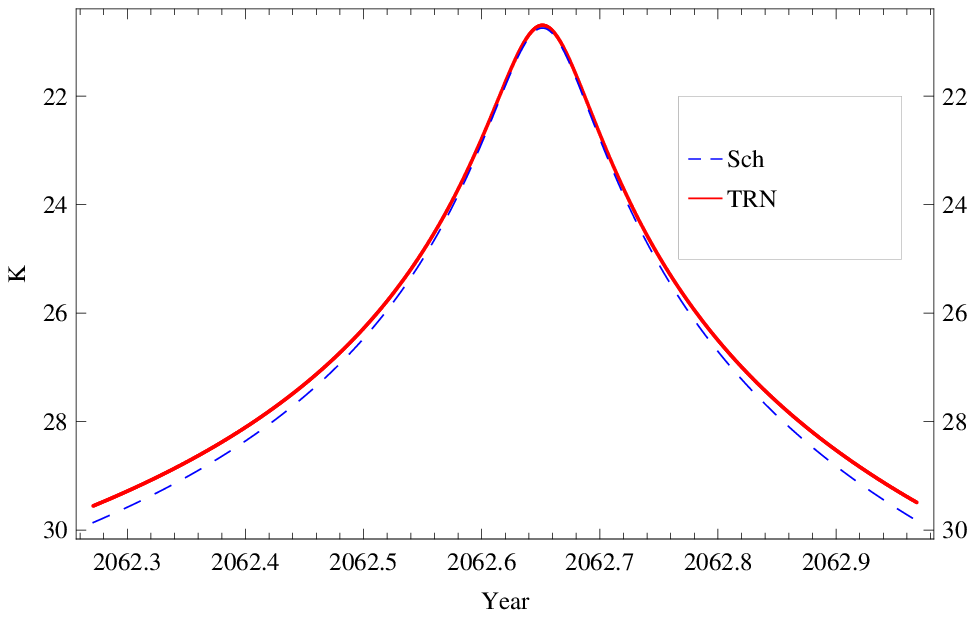}
 \caption{The light curve for the secondary image of S14 near its periapse and peak brightness (top) shows a small effect due to a $\frac{1}{r^2}$ term at peak. The effect for S6 around its peak brightness (bottom) is even smaller. In both cases, $q= -1.6$ for the TRN curve.}
 \label{fig:s14curve}
 \end{center}
 \end{figure}

At  peak brightness, there is a difference of about 0.4 magnitudes between the secondary image in a Schwarzschild spacetime and the image with $q= -1.6$. The image is 44\% brighter with $q=-1.6$ at the periapse of S2, which will  next take place in 2018. The observational prospects of this difference will be discussed in Sec. \ref{sec:discussion}. However, for the extremal black hole with $q= 0.25$, the image is 0.03 magnitudes fainter than for the Schwarzschild spacetime, a distinction that is beyond projected observational capacity.  If the metric around the black hole has 5-dimensional behavior encoded in a $\frac{1}{r^2}$ term in the metric, than for a large enough value of $-q$, there is a significant variation in the brightness of the secondary image. A similar analysis around peak brightness is performed for the stars S14 and S6 as well. As can be seen in Fig. \ref{fig:s14curve}, the effect of the $\frac{1}{r^2}$ term is not as pronounced for these stars. Consequently, we have suppressed the curve for the extremal RN spacetime, as it shows little difference from the Schwarzschild value near peak.

The explanation is the alignment of each star with Sgr A* at periapse. S2, because of the inclination of its orbit with respect to the plane containing Earth and Sgr A*, is not well aligned with the optic axis at the time of peak brightness ($\gamma_0 = 45.4 ^\circ$). Despite this, the secondary image is still bright because of the close approach of S2 to Sgr A*. However, because the source is at a large angle, the secondary image appears very close to the black hole (43 $\mu$as  in the Schwarzschild, $q=0$ limit). Because light passes so close to Sgr A*, as close as 3.5 Schwarzschild radii, the $\frac{Q}{r^2}$ term is more dominant and its effects are more pronounced. 

\begin{figure}[h]
 \begin{center}
\includegraphics[width=0.5 \textwidth]{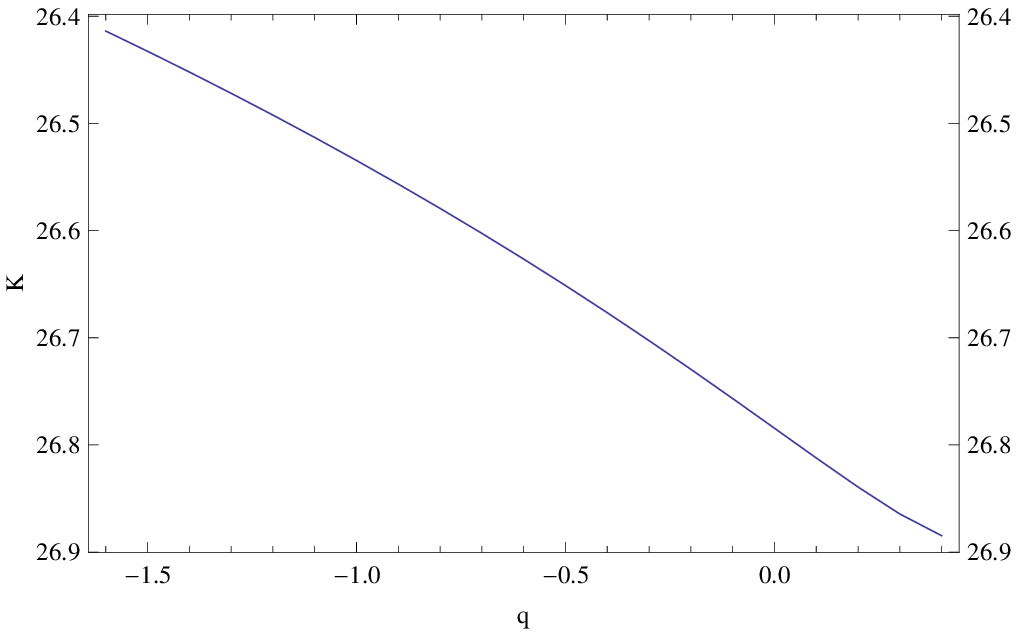}
\includegraphics[width=0.5 \textwidth]{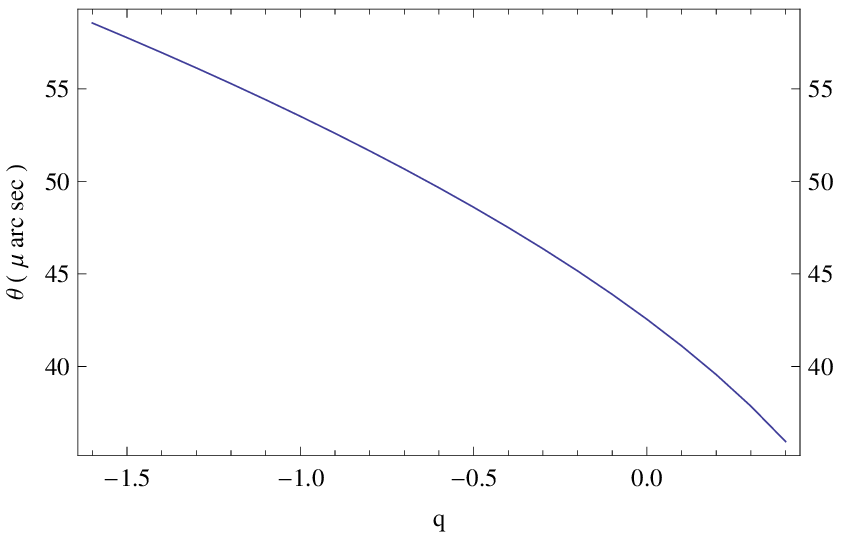}
 \caption{This figure displays the relationship between $q$ and image observables for S2. The relationship between $q$ and the apparent magnitude in the K-band (top) shows that brightness is inversely related to $q$ . This is related to the fact that the angular position of the secondary image, $\theta$, also has an inverse relationship with $q$ (bottom). }
 \label{fig:s2q}
 \end{center}
 \end{figure}

On the other hand, S14 is more closely aligned with the optic axis at periapse and peak brightness ($\gamma_0 = 9.5 ^\circ$). Hence, the secondary image, while brighter relative to the source than S2's image, is further away from the optic axis (136 $\mu$as) and the point of closest approach is further (12.8 Schwarzschild radii). Hence, the effects of the $\frac{1}{r^2}$ term are less noticeable than in the case of S2. S6 is highly aligned with the optic axis ($\gamma_0 = 3.6 ^\circ$) and, correspondingly, the null geodesic forming the secondary image passes even further from the black hole, giving an image position of 316 $\mu$as \cite{bozzamancini2009} and a point of closest approach of 30.5 Schwarzschild radii. In Fig. \ref{fig:s2q}, we examine the relationship between the $q$ parameter and observables such as image magnitude and position for the secondary image of S2 at periapse. 
The relationship between $q$ and observables is very similar for the secondary images of S6 and S14, so we have not shown graphs illustrating these relationships. Instead, we have compiled Tables \ref{table:magq} and \ref{table:magtheta}, which explore the variation of image observables with tidal charge parameter $q$ for both positive and negative values of $q$. We have explored values of $q$ as high as $0.4$ which represents a super-extremal RN black hole. 

\begin{table*}[t]
\begin{tabular}{c| c c c c c c c c c c c}
\hline
\hline
  Star & $\frac{2}{5}$& $\frac{1}{5}$ & 0& $-\frac{1}{5}$ & $-\frac{2}{5}$& $-\frac{3}{5}$ & $-\frac{4}{5}$ & $-1$ & $-\frac{6}{5}$& $-\frac{7}{5}$ & $-\frac{8}{5}$ \\
\hline
S2  & 26.88& 26.83& 26.78 & 26.73 & 26.68 & 26.63 & 26.58 & 26.53& 26.49 & 26.45& 26.41  \\
S14 & 23.54 & 23.52 &23.5 & 23.48& 23.47& 23.45 & 23.43& 23.42& 23.40& 23.39 & 23.37   \\
S6 & 20.76 & 20.75& 20.74& 20.74& 20.73& 20.72& 20.72& 20.71& 20.71& 20.70& 20.69\\
\hline
\end{tabular}
\caption{\label{table:magq} This table gives the peak brightness for each star at several values of $q$. As $|q|$ gets bigger, the  effect gets larger in all cases. However, the increase in brightness with the increase in $|q|$ is more pronounced for stars that are not aligned with the optic axis. For S2, additional tidal charge makes a significant increase in the brightness of the secondary image. Since only a small value is allowed for a positive $q$ (to avoid a naked singularity) and $q$ has few bounds in the negative direction, negative $q$ is a more promising avenue for exploration.}
\end{table*}

The primary contribution to the differences in the image magnifications is the tangential magnification
\begin{equation}
\mu_t= \frac{\sin \theta}{\sin \gamma}.
\end{equation}
as the image of the position $\theta$ grows larger as $-q$ does. This makes $\theta$ larger in a TRN spacetime (for negative $q$) and, therefore, $\mu_t$ larger as well. When $r_0$ is small, the contribution of the $\frac{1}{r^2}$ term is more important. 

\begin{table*}[t]
\begin{tabular}{c| c c c c c c c c c c c}
\hline
\hline
  Star & $\frac{2}{5}$ & $\frac{1}{5}$ & 0& $-\frac{1}{5}$ & $-\frac{2}{5}$& $-\frac{3}{5}$ & $-\frac{4}{5}$ & $-1$ & $-\frac{6}{5}$& $-\frac{7}{5}$ & $-\frac{8}{5}$ \\
\hline
S2 &35.9 & 39.6 & 42.6 & 45.2 & 47.5 & 49.7 & 51.6 & 53.5& 55.3 & 57.0& 58.6  \\
S14 & 130.5& 133.1& 135.6 & 138.0& 140.3& 142.5& 144.7& 146.8& 148.8& 150.8& 152.8\\
S6 & 311.5 & 313.8 &316.1 & 318.3& 320.6& 322.7 & 324.9& 327.0& 329.1& 331.2 & 333.2   \\
\hline
\end{tabular}
\caption{\label{table:magtheta} This table gives the angular position in $\mu$as of the secondary image relative to the optic axis at the star's periapse. As $|q|$ gets bigger, the effect gets larger in all cases.However, for positive $q$, $\theta$ gets smaller as $q$ gets bigger. $\theta$ gets larger with negative $q$. Although the increase in angular position with increasing tidal charge seems to be very similar in the case of all the stars, the shift represents a bigger relative shift for the ones with smaller angular position.}
\end{table*}

The effect of black hole spin on the secondary images was studied by \cite{bozzas2} and commented on by \cite{bozzas2revised}. In this paper, we have not studied the effects of the Kerr metric for several reasons. Firstly, evaluation of lensing in a Kerr spacetime is far more challenging than the study of a TRN spacetime, so we have started with a less ambitious project. Also, while for higher-order images, spin can greatly change the magnification, this is not true for secondary images \cite{bozzas2, bozzas2revised}, and the effect of the black hole's spin is less significant. Still, the magnification of secondary images depends on the magnitude and direction of the black hole's spin and secondary images in the Kerr spacetime are worthy of further study. Some work on rotating black holes and their gravitational lensing effects has already been done \cite{bozzakerr, bozzakerr2}, including some work on rotating braneworld black holes \cite{aliev}. 

\section{\label{sec:discussion} Discussion}

Of the three stars studied in this paper, S14 and S6 have brighter secondary images at periapse compared to S2's image because of the more edge-on nature of their orbits relative to our line of sight with Sgr A*. While the orbit of S2 is not aligned close to edge-on, its periapse is the closest amongst known stars, causing its secondary image to be very bright. In addition, S2 will next be at periapse in 2018, allowing for a more immediate study of its image's properties. 

As mentioned above, the brightness of secondary images is dependent on $q$ because the images are either pushed closer or further from the optic axis when there is a $\frac{1}{r^2}$ term in the metric. This is directly related to the difference in the size of the event horizon and photon sphere due to the value of $q$. Another, and perhaps easier, way of determining the metric around the black hole would be to measure the shape and size of the black hole's horizon or photon sphere. A preliminary attempt at this has been made \cite{NatureHorizon}, but they were unable to identify the observed structure with the black hole itself. At present, there are no constraints which would prevent the size of the event horizon from being significantly larger or smaller than predicted in a Schwarzschild spacetime (but not by an order of magnitude \cite{flaring}). In addition, since the photon sphere is so small, even a relatively large percentage change in its size corresponds to only a few $\mu$as. Resolving the difference between two proposed photon sphere sizes may be beyond the capabilities of projected future instruments. In this case, observing secondary images may provide more information. 

There has been some discussion about observing these secondary images with the upcoming generation of telescopes \cite{bozzamancini2009}. A very promising project is the MCAO Imaging Camera for Deep Observations (MICADO) telescope \cite{micado} at E-ELT. It is expected to be able resolve images as faint as $m_K=30.1$ and will have a resolution of up to 6 mas in imaging mode (10 mas in the K-band) and an astrometric accuracy of up 10 $\mu$as. It also will have a photometric accuracy of 0.03 magnitudes. From the data shown above, trying to resolve the difference between an image's position in the Schwarzschild metric and an image's position in the TRN metric is next to impossible given the small separation between these positions. 

Therefore, it is important to explore not only image (and photon sphere) positions, but to use image magnfications as a complementary avenue for exploring the metric. If the secondary images in this paper were isolated point sources, MICADO would not have any problems detecting them and even differentiating between two predicted values for image brightness (assuming the difference is larger than the photometric accuracy). However, these images will be very close to and essentially unresolvable from Sgr A* and its crowded environment. This is not a fatal flaw, because Sgr A* is very faint in the Near Infared K-band ($\lambda = 2.2 \mu$m), and it may be feasible to subtract out the quiescient state of Sgr A* in the K-band. Some studies \cite{flares, flares2, ghezconstant, constant2} claim that Sgr A* has a highly variable ``quiescient" state of $m_K \approx 17$. In addition, there are occasional flares that can be brighter than $m_K = 16$ and last on a scale of hours. While the flares are thought to originate very close to the black hole, it is not clear whether the ``quiescient" radiation in the K-band comes from Sgr A* itself or whether it comes from the fact that the lower resolution (65 mas) in the survey includes one or more sources in the area near Sgr A*. The currently known star with the closest approach is S2, which appears to be 11 mas from Sgr A* at the point of closest approach- at all points, MICADO should be able to resolve S2 and many other closely moving stars from Sgr A*. There may be other, unidentified stars that are currently conflated with radiation from Sgr A* but will be separately with MICADO's increased resolution. Better resolution, combined with further understanding of the K Luminosity Function near the Galactic center means that observations of secondary images is a real possibility. MICADO's photometric accuracy should be able to distinguish many of the image brightness differences discussed in this paper. This assumes that when viewed with a fine enough resolution, Sgr A* does not emit too brightly in the K-band and source crowding is not insurmountable. Even if flares are observed, they persist for a time scale much shorter than secondary images (which remain bright for months) and should be easily distinguished from the quiescient state. Additionally, it is expected that an additional $10^0$-$10^2$ S stars will be found in the central milli-parsec of the galaxy \cite{milliparsec}, yielding additional and perhaps even better candidates to observe secondary images with the right properties to give us insight into the metric near the black hole. It may not be possible to accurately treat stars that close to Sgr A* with the thin-lens approximation, but in that case, a more exact numerical treatment can be utilized \cite{bozzas2, cunningham}.

Observing secondary images of S stars will be challenging, but we may very well find that it is possible to observe faint secondary images and use their properties to give us insight into the true nature of gravity. This exciting prospect should be an additional motivation for the next generation of observational instruments aimed at the Galactic center.

\section{Acknowledgements}

I would like to thank R.K. Sheth and J. Khoury for their unwavering support of my research, R. Davies for very helpful discussions related to the observational aspects of this paper, and G. Bernstein, S. Doelman, T. Pumard, A. Lidz, B. Jain, A. Ghez, B. Greenbaum, and F. Melia for helpful conversations.

\bibliographystyle{h-physrev}

\end{document}